\documentclass[10pt,conference]{IEEEtran}
\IEEEoverridecommandlockouts
\usepackage{cite}
\usepackage{amsmath,amssymb,amsfonts}
\usepackage{algorithmic}
\usepackage{graphicx}
\usepackage{textcomp}
\usepackage{xcolor}
\def\BibTeX{{\rm B\kern-.05em{\sc i\kern-.025em b}\kern-.08em
T\kern-.1667em\lower.7ex\hbox{E}\kern-.125emX}}

\usepackage{newfloat}
\usepackage{listings}

\lstdefinelanguage{stripspddl}
{
  sensitive=false,
  morecomment=[l]{;}, 
  alsoletter={\#,:},   
  morekeywords={not, and, :action, :parameters, :precondition, :effect, :constants, :init, :goal, :types, :predicates, :requirements, define}
}

\usepackage{hyperref}

\begin{document}

\title{Optimal Layout Synthesis for Quantum Circuits \\
as Classical Planning (full version)%
\thanks{This is the full technical report for
I. Shaik and J. van de Pol, {\em Optimal Layout Synthesis for Quantum Circuits
as Classical Planning}. In: Proc. IEEE/ACM IC on Computer-Aided Design, 
(ICCAD'23), San Francisco, California, USA, 2023.}}

\author{\IEEEauthorblockN{Irfansha Shaik}
\IEEEauthorblockA{Department of Computer Science \\
Aarhus University\\
Aarhus, Denmark\\
\href{mailto:Irfansha.Shaik@cs.au.dk}{Irfansha.Shaik@cs.au.dk}}
\and
\IEEEauthorblockN{Jaco van de Pol}
\IEEEauthorblockA{Department of Computer Science \\
Aarhus University\\
Aarhus, Denmark\\
\href{mailto:Jaco@cs.au.dk}{Jaco@cs.au.dk}}
}

\maketitle
\thispagestyle{plain}
\pagestyle{plain}

\begin{abstract}
In Layout Synthesis, the logical qubits of a quantum circuit are mapped to the physical qubits
of a given quantum hardware platform, taking into account the connectivity of physical qubits.
This involves inserting SWAP gates before an operation is applied on distant qubits.
Optimal Layout Synthesis is crucial for practical Quantum Computing on current error-prone hardware:
Minimizing the number of SWAP gates directly mitigates the error rates when running quantum circuits.

In recent years, several approaches have been proposed for minimizing the required SWAP insertions.
The proposed exact approaches can only scale to a small number of qubits.
Proving that a number of swap insertions is optimal is much harder than producing near optimal mappings.

In this paper, we provide two encodings for Optimal Layout Synthesis as a classical planning problem.
We use optimal classical planners to synthesize the optimal layout for a standard set of benchmarks.
Our results show the scalability of our approach compared to previous leading approaches.
We can optimally map circuits with 9 qubits onto a 14 qubit platform, which could not be handled before
by exact methods.
\end{abstract}


\section{Introduction}
\label{sec:Introduction}

Quantum algorithms can speed up certain computational problems. For example,
Shor's Algorithm \cite{DBLP:conf/focs/Shor94} provides an exponential speed-up for factorization,
and Grover's algorithm \cite{DBLP:conf/stoc/Grover96} provides a quadratic speed-up for database search.
In recent years, quantum computing is envisaged as a way to solve hard problems \cite{arute2019quantum},
out of reach for classical computers.
Formulating new algorithms to solve more problems faster on quantum computers, 
and enabling compilation of those algorithms on actual quantum hardware is of great interest.

In this paper, we focus on the latter part, especially mapping a quantum circuit to
an arbitrary quantum architecture. Following \cite{DBLP:journals/tc/TanC21}, we distinguish
logical synthesis and layout synthesis for quantum circuits. \emph{Logical synthesis} constructs
an (efficient) quantum circuit of quantum gates on a number of logical qubits. 
Following \cite{DBLP:journals/tcad/AmyMMR13,cross2017open,EQGR2011Miller}, 
arbitrary quantum circuits can be represented by a few simple gates.
We assume that the resulting quantum circuit consists
of any standard unary gates and binary Controlled-NOT (CNOT) gates.
A CNOT gate acts on two qubits. If the control qubit is true, then the target qubit is flipped.
After logical synthesis, the gates and the order of their application to the
logical qubits is fixed. 

We focus on \emph{Layout synthesis}, mapping the logical qubits onto the qubits of some
physical hardware platform.
High level algorithms assume that every pair of (logical) qubits is connected.
However, in actual hardware platforms, often only a subset of (physical) qubit connections are realized.
So one needs to compile the logical circuit to a physical hardware
platform, such that every CNOT gate is applied only on neighboring qubits.
It is not always possible to find a fixed mapping from logical qubits to physical qubits,
such that every CNOT is applied on neighbors. In such cases, SWAP gates can be inserted,
to swap the values of two connected qubits. A SWAP gate can be built from 3 CNOT gates.
The goal of layout synthesis is to insert a minimal number of SWAP gates into the circuit,
to move logical qubits to physical qubits, satisfying all neighbor constraints on 
subsequent CNOT gates.

\paragraph*{State-of-the-Art and Related Work}
Layout Synthesis has many names, for example, Transpilation, Compilation, Routing problem etc.
An extensive overview and comparison of existing approaches is provided in \cite{DBLP:journals/tc/TanC21},
discussing both exact and heuristic approaches.
Several heuristic approaches exist, for instance based on Dynamic Programming \cite{DBLP:conf/cgo/SiraichiSCP18},
A* search \cite{DBLP:journals/tcad/ZulehnerPW19}, and Temporal Planning \cite{DBLP:conf/ijcai/VenturelliDRF17,DBLP:conf/aips/BoothDBRVF18,venturelli2019quantum,DBLP:conf/ecai/DoWOVRF20}.
Concrete tools include Qiskit's SABRE \cite{DBLP:conf/asplos/LiDX19} and
Hardware-Aware (HA) \cite{Niu2020ha}.
While the state-of-the-art heuristic approaches are fast and scalable,
it has been observed that the solutions are non-optimal~\cite{DBLP:conf/dac/WilleBZ19}.
In fact, \cite{DBLP:journals/tc/TanC21} observed large optimality gaps,
up to 1.5-12x and 5-45x on average on major quantum computing platforms.

For practical quantum computing on current NISQ machinery (noisy, intermediate scale quantum computers),
the primary goal is to reduce the error rate.
Since with each new gate the error rate increases, it is essential to minimize
the number of additional SWAP gates in the Layout Synthesis. Therefore, in this
paper we will focus on exact approaches.

Recently, several exact approaches have been proposed for optimal Layout Synthesis.
An exact approach based on Dynamic Programming was proposed in \cite{DBLP:conf/aspdac/ItokoRIMC19},
but it scales only to small circuits.
Layout Synthesis was formulated as an SMT encoding in \cite{DBLP:conf/dac/WilleBZ19},
where propositional constraints are given for satisfying connectivity and integer costs are assigned
to additional swap gates.
The authors report improved optimal plans compared to heuristic approaches.
They use an exponential number of variables to represent all permutations for the initial mapping.
In a followup work~\cite{10.1145/3593594}, optimal subarchitectures are considered
to reduce the number of permutations by mapping to a subset of physical qubits.
In \cite{DBLP:conf/iccad/TanC20}, the tool OLSQ is presented, 
based on an improved SMT encoding of permutations using a polynomial number of variables
in the number of qubits.
Those authors also allow optimizing error rates when individual error rates are given for qubits and gates.
In \cite{DBLP:conf/micro/MolaviXDPTA22}, it is proposed to solve Layout Synthesis as a MAXSAT problem.

In theory, both SMT and MAXSAT approaches can prove lower bounds on the number of SWAP gates, 
thus guaranteeing optimality.
However, optimal Layout Synthesis is an NP-complete problem \cite{DBLP:conf/socs/BoteaK018}.
In practice, most tools make some approximations.
For instance, they limit the number of added swap gates, or they
switch to heuristic search after some time limit.
For example, the tool \texttt{satmap} by \cite{DBLP:conf/micro/MolaviXDPTA22} only produces near optimal solutions,
due to a restriction on the number of swaps.
Even with these restrictions, all three approaches report improved optimality compared to heuristic approaches.
However, the improved (near) optimality comes at the cost of time and memory.
We aim at improving the performance, while preserving the optimality of the exact approaches.

\paragraph*{Contribution}
In this paper, we provide the first encoding of Optimal Layout Synthesis 
as a classical planning problem. The possible actions in this planning problem are: 
1) to map a logical qubit to an initial physical qubit;
2) to apply a CNOT gate to the proper qubits; or 3) to insert an additional SWAP gate.
In the latter case, we allow swapping logical qubits, as well as swapping a logical
qubit with an extra ancillary qubit, since this can lead to even shorter plans.
The conditions on the actions ensure that the solution satisfies all constraints.
The shortest plan corresponds to a layout with the minimal number of SWAP gates.
This enables the use of optimal planning tools to find the optimal layout.

In fact, we provide two encodings. The first one is based on \emph{global levels} to resolve dependencies. Viewing these levels
as discrete time stamps, this encoding
is close to the temporal planning approach~\cite{DBLP:conf/ijcai/VenturelliDRF17,DBLP:conf/aips/BoothDBRVF18,venturelli2019quantum,DBLP:conf/ecai/DoWOVRF20}, 
but we use the simpler fragment
of classical planning, and moreover our encoding is exact, while their approach minimizes
makespan, but not necessarily the number of SWAP gates.
Our second, more efficient but still exact encoding, avoids the explicit representation of time steps
and is based on the relative order induced by \emph{local dependencies}
between CNOT gates.

We implement both encodings in the open source translator Q-Synth
(\url{https://github.com/irfansha/Q-Synth/}), written in Python. 
It takes two inputs: a quantum circuit with unary and binary CNOT gates in the standard 
OPENQASM 2.0 format \cite{cross2017open},
and a directed graph, modeling the connectivity graph of a hardware architecture platform. 
We use Qiskit libraries~\cite{Qiskit} for parsing, printing and extracting layers from a given circuit.
Initially, we ignore unary gates and compute the dependencies between the CNOT gates.
The output of the translator is a planning problem 
represented in PDDL (Planning Domain Definition Language)~\cite{DBLP:journals/aim/McDermott00}.
This enables to use off-the-shelf classical planners
for computing optimal plans for these PDDL instances.
The optimal plan is translated back to an OPENQASM circuit
by re-inserting the unary gates.

We experimented with the planner Fast Downward~\cite{helmert2011fast} 
(with multiple backends~\cite{domshlak2011bjolp, sievers2018fast}) and
with Madagascar~\cite{Rintanen2014MadagascarS}, which is based on an encoding into SAT
(propositional satisfiability).
Other planners, like those based on QBF (Quantified Boolean Logic)~\cite{DBLP:conf/aips/ShaikP22}
could be used as well.
We report the performance of our two approaches on some standard benchmarks and
compare them to a number of existing exact 
(QMAP~\cite{DBLP:conf/aspdac/BurgholzerSW22, 10.1145/3593594} and OLSQ~\cite{DBLP:conf/iccad/TanC20}) 
and heuristic approaches (QISKIT's SABRE~\cite{DBLP:conf/asplos/LiDX19}). 
As reported before, the heuristic approach SABRE is very fast, but often uses more SWAP gates than necessary.
In all solved cases, we computed the correct minimal number of required SWAP gates.

We demonstrate that the performance of our approach is better than previous exact approaches.
Moreover, our Local encoding performs much better than our Global encoding.
The existing tool QMAP performed better than OLSQ in our experiments on a 5 qubit platform.
On the other hand, OLSQ performed better than QMAP on a 14 qubit platform.
Our Local approach is up to two orders of magnitude faster than OLSQ and 
solves 8 unique instances that OLSQ could not solve in 3 hours.
Our approach finds optimal mappings of 9-qubit circuits onto a 14-qubit platform,
where previous exact tools, OLSQ and QMAP, could only map 5-qubit and 4 qubits circuits respectively on the same platform.

\section{Preliminaries}
\label{sec:Preliminaries}

\subsection{Layout Synthesis for Quantum Circuits}
\label{subsec:layoutsynthesis}

Layout Synthesis is mapping a logical circuit to some physical quantum computer architecture.
Higher level algorithms make some assumptions, for example one-to-one connectivity is assumed among all qubits.
However, physical architectures can have many restrictions, for example limited connectivity between qubits
where binary gates can be applied.
Particular qubits could also have different error rates and time durations.
One takes into account such considerations when mapping to some architecture.
In case no mapping exists such that the logical circuit can be mapped, SWAP gates can be added between neighbors
to move the logical qubits to connected qubits.
Note that using ancillary qubits can reduce the number of required SWAP gates.
In this work, we focus on minimizing the number of SWAP gates
and consider the use of planning for error rate optimization as future work.

\begin{figure}[b]
  \centering
  \includegraphics[scale=0.3]{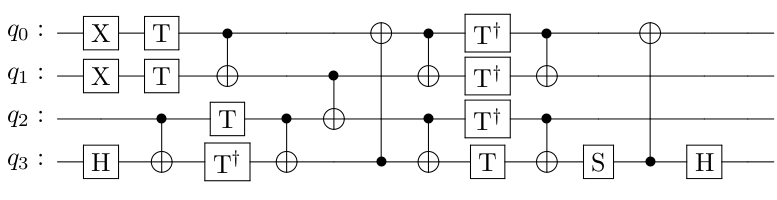}
  \caption{Adder circuit before mapping to IBM-QX2.}
  \label{fig:adder}
  \end{figure}

\begin{figure}[htbp]
    \begin{minipage}{0.55\columnwidth}
    \begin{small}
    \begin{verbatim}
    OPENQASM 2.0;
    include "qelib1.inc";
    qreg q[4];
    x q[0];//g1
    x q[1];//g2
    h q[3];//g3
    cx q[2], q[3]; //g4
    t q[0];//g5
    t q[1];//g6
    t q[2];//g7
    tdg q[3];  //g8
    cx q[0], q[1]; //g9
    cx q[2], q[3]; //g10
    cx q[3], q[0]; //g11
    cx q[1], q[2]; //g12
    cx q[0], q[1]; //g13
    cx q[2], q[3]; //g14
    tdg q[0];  //g15
    tdg q[1];  //g16
    tdg q[2];  //g17
    t q[3];//g18
    cx q[0], q[1]; //g19
    cx q[2], q[3]; //g20
    s q[3];//g21
    cx q[3], q[0]; //g22
    h q[3];//g23
    \end{verbatim}
  \end{small}
    \end{minipage}\hfil
    \begin{minipage}{0.4\columnwidth}
      \includegraphics[scale=0.22]{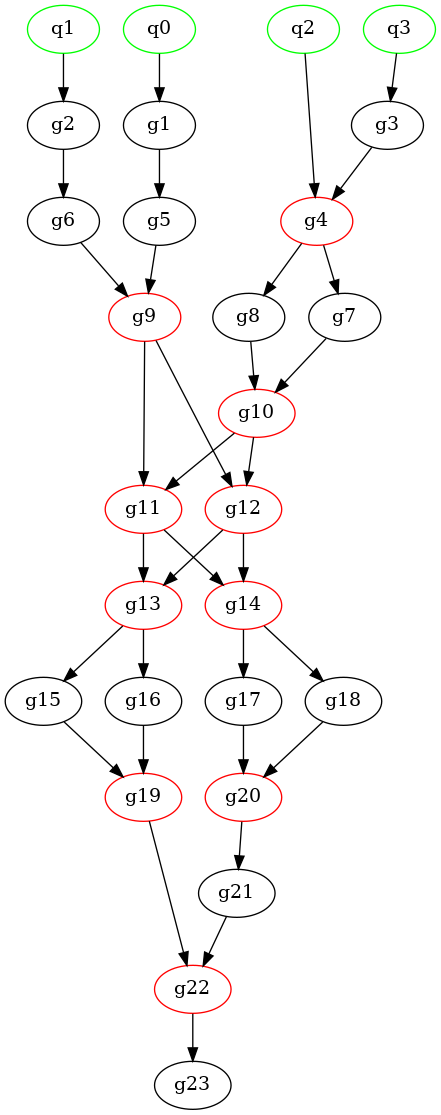}
    \end{minipage}
    \caption{Adder circuit in OPENQASM format and in DAG format,
    showing the gate dependencies.}
    \label{fig:addercircuit}
    \end{figure}

We illustrate layout synthesis on an adder circuit, a standard example from \cite{DBLP:conf/iccad/TanC20}.
The traditional circuit representation of the circuit is shown in Fig.~\ref{fig:adder}.
It contains 10 binary CNOT gates and 13 unary gates.
Fig.~\ref{fig:addercircuit} shows the adder circuit in OPENQASM format (with gate numbers as comments).

\begin{figure}[b]
\centering
\includegraphics[scale=0.24]{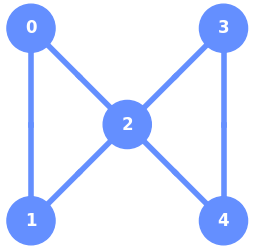} 
\caption{Coupling map for IBM-QX2 (Tenerife)}
\label{fig:couplingmapibmqx}
\end{figure}

\begin{figure}[b]
\centering
\includegraphics[scale=0.3]{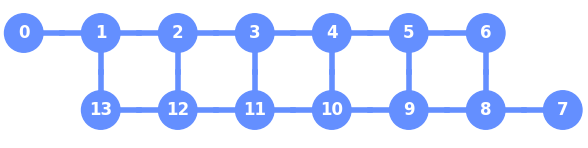}
\caption{Coupling map for IBM Melbourne}
\label{fig:couplingmapmel}
\end{figure}

\begin{figure*}[htbp]
\centering
\includegraphics[scale=0.3]{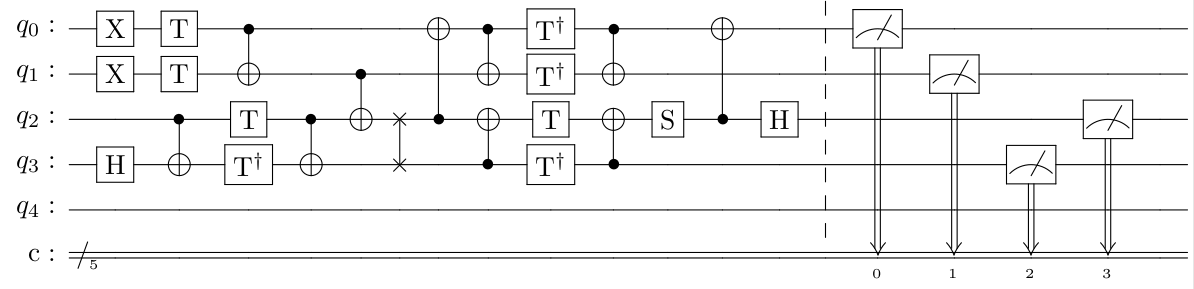}
\caption{Adder circuit after mapping to IBM-QX2, with 1 optimal swap insertion.
Measurement gates indicate the final location of the logical qubits.}
\label{fig:addermapped}
 \end{figure*}

Figure \ref{fig:addercircuit} also provides a DAG representation with dependencies between gates, 
as computed by Qiskit \cite{Qiskit}.
The green nodes (inputs) are logical qubits, whereas the red nodes represent the (binary) CNOT gates, 
and the other nodes are unary gates.
We use the gate numbers to represent the dependencies between gates in the DAG.
There are 11 layers in total, if we do not count the top qubit layer (the inputs are not part of the circuit).
All gates in the same layer act on independent qubits, so they can be executed independently.
We use this fact in Section \ref{subsec:globallevelbased} to handle dependencies between CNOT gates.

Figure \ref{fig:couplingmapibmqx} illustrates the IBM platform IBM-QX2 with 5 physical qubits.
In general, the connections between qubits are directed, and our approach can handle
directed coupling graphs. However, other exact tools can only handle symmetric connections,
so for a fair comparison we only consider bidirectional platforms in this paper.

To map the adder circuit onto any platform, one needs 4 physical qubits connected as a square
since it has CNOT gates on the qubit pairs $(q0,q1),(q1,q2),(q2,q3),(q3,q0)$.
However, IBM-QX2 has no 4 qubits that are connected in the form of a square, thus SWAP gates are needed.
We observe from our results in Table \ref{table:tenerifedata} that 1 SWAP insertion is sufficient.
The circuit in Figure \ref{fig:addermapped} shows the mapped circuit with 1 additional SWAP gate.
First, the logical qubits ${q0,q1,q2,q3}$ of the circuit are mapped to physical qubits ${q0,q1,q2,q3}$.
For the mapped circuit, the order of gates g11 and g12 is reversed,
and a SWAP gate on qubits q2 and q3 is added in between.
Since g11 and g12 are in the same layer, they are independent, and this change preserves the correctness.
All gates after g11 and g12 must be mapped to the swapped physical qubits:
Essentially, from this point on, q2 and q3 are interchanged.

The Melbourne platform (Fig.~\ref{fig:couplingmapmel}) contains a subset of 4 qubits that form a square.
Consequently, the adder circuit does not require any additional SWAP gates when mapped
on Melbourne.
We can indeed find a mapping without any SWAP gates as shown in Table \ref{table:melbournedata}.

\subsection{Classical Planning}
\label{subsec:classicalplanning}

One of the main applications of Automated Planning \cite{ghallab2004automated} is scheduling.
In Section \ref{sec:layoutsyntheisaspp}, we will model Layout Synthesis as a planning problem.
Classical Planning is finding a valid sequence of deterministic actions from a single initial state to some goal state, 
where the initial state and the effect of all actions are completely known.
The \emph{Planning Domain Definition Language} (PDDL)~\cite{DBLP:journals/aim/McDermott00} 
is a standard domain-specific language for classical planning problems used in International Planning Competitions (IPC).
As shown in Listing \ref{lst:domainstructure}:
The domain file specifies types, predicates and actions.
For predicates one can optionally specify types, as we will later see in the Section \ref{sec:layoutsyntheisaspp}.
PDDL is an action-centered language; the actions essentially specify how the world changes by preconditions and effects.
Action conditions and effects are represented similar to First-Order-Logic, using parameters to achieve compact descriptions. 
An example for an action is \texttt{map\_initial} with parameters \texttt{?l} and \texttt{?p} in Section \ref{subsec:globallevelbased}.
The problem file specifies objects, initial state and goal condition%
as shown in Listing \ref{lst:problemstructure}
Initial and Goal states are represented using a set of (negated) propositions.

Given a planning problem in PDDL format, one can use off-the-shelf planners such as Fast Downward, Madagascar, etc.
A planning problem description can also assign costs to actions.
In Planning competitions, planners are categorized based on how they handle these costs.
Optimal Planners return only plans with optimal cost;
Satisfying Planners, on the other hand, return some plan quickly, regardless of the cost.
Nevertheless, given enough time, some satisfying planners such as Fast Downward Stone Soup (FDSS)~\cite{helmert2011fast} can optimize the cost to a large extent.
Note that if cost is not specified, each action is considered to have cost 1.

\begin{lstlisting}[caption={Domain file structure},label={lst:domainstructure},
language=stripspddl,basicstyle=\small,mathescape]
  (define (domain Quantum)
   (:requirements :typing)
   (:types ...)
   (:predicates ...)
   (:action ...)...)
\end{lstlisting}

\begin{lstlisting}[caption={Problem file structure},label={lst:problemstructure},
  language=stripspddl,basicstyle=\small,mathescape]
  (define (problem example)
   (:domain Quantum)
   (:objects ...)
   (:init ...)
   (:goal (and ...)))
\end{lstlisting}

\section{Layout Synthesis as Classical Planning}
\label{sec:layoutsyntheisaspp}

In this section, we describe two encodings of layout synthesis as a classical planning problem,
represented in PDDL.%
\footnote{Our Quantum Circuit Layout domain received the Outstanding Domain Submission Award at the 
International Planning Competition IPC 2023.}
Both encodings include the following actions:
\begin{itemize}
\item \texttt{apply\_cnot}, to apply the CNOT gate of two logical qubits on two physical qubits.
\item \texttt{swap}, to insert a SWAP gate. We allow swapping
two mapped logical qubits, as well as swapping
a mapped logical qubit with an ancillary qubit.
\end{itemize}
Both encodings include the following predicates, to model the state after a partial plan:
\begin{itemize}
  \item \texttt{mapped(?l,?p)}, to keep track on which physical qubit a logical qubit is currently mapped.
  \item \texttt{rcnot(...)}, keeping track which CNOT gates are still required
  (all CNOT gates are still required in the initial state, none should be required in the goal state).
\end{itemize}
The planning problem defines the following constraints on valid plans:
\begin{itemize}
  \item Initial Mapping: The initial mapping from logical to physical qubits must be injective.
  \item Valid Gate Dependencies: Gates must be applied to the correct qubits in proper order.
  \item Valid CNOT and SWAP gates:
      All CNOT and SWAP gates are only applied on neighboring physical qubits.
\end{itemize}
As a consequence, since all CNOT gates must be applied anyway, the shortest valid plan
will correspond to the correct layout, with the minimal number of SWAP gates inserted.

Since the hardware platform only constrains 2-qubit gate connectivity, layout synthesis can focus 
on the CNOT gates and their dependencies only. The unary gates can be reinserted later.
Ignoring the unary gates, every CNOT gate has at most two direct dependencies. 
We illustrate this again on the DAG of the adder-circuit of Fig.~\ref{fig:addercircuit}.
There, $g4$ depends
on the input gates $g2$ and $g3$; gate $g10$ only depends on $g4$; and gate $g11$ depends on the CNOT gates
$g9$ and $g10$. In general, a CNOT gate could also depend on one input qubit and one CNOT gate, but
this does not happen in the example of Fig.~\ref{fig:addercircuit}.

We parse the circuit and extract the CNOT gates with their dependencies.
After finding a solution, we reconstruct the circuit including single qubit gates.
We first encode $n$ logical qubits as objects $\{l_0,\dots,l_{n-1}\}$ of type \texttt{lqubit},
and $m$ physical qubits using objects as $\{p_0,\dots,p_{m-1}\}$ of type \texttt{pqubit}.
In our experiments, we do not specify costs, so each additional SWAP gate increases the cost by 1.
Alternatively, one could specify an explicit cost for SWAP gates; 
for example a cost of 3, representing 3 underlying CNOT gates.

The first encoding (Section \ref{subsec:globallevelbased}) is based on 
\emph{global levels}, corresponding to the layers with CNOT gates in the dependency graph. 
This encoding uses two extra actions:
\texttt{map\_initial}, mapping a logical qubit to a physical qubit in the first time slice,
and action \texttt{move\_depth}, to advance to the next time slice.
The second, more efficient encoding (Section \ref{subsec:localdependencybased}), avoids the explicit representation of time steps
and is based on the relative order of \emph{local dependencies} between CNOT gates (and logical input qubits).

\subsection{Global Level Based Model}
\label{subsec:globallevelbased}

\subsubsection{Domain Specification}
\label{subsubsec:domainspec}
In this section, we specify the predicates and actions required for the domain specification.
Note that a domain file specifies the predicates and the actions of the problem domain.
The predicates describe the state of the world, whereas the actions describe the change from one state to another.

\paragraph*{Valid Initial Mapping}
For modelling the first constraint i.e., mapping logical qubits to physical qubits, we define three predicates and one action.
The predicate \texttt{mapped} specifies that the logical qubit \texttt{?l} is mapped to the physical qubit \texttt{?p}.
The two auxiliary predicates \texttt{mapped\_pq}/\texttt{mapped\_lq} describe if a physical/logical qubit
is already mapped, since one needs an injective mapping.
The parameter \texttt{?p} (\texttt{?l}) in these predicates is of type \texttt{pqubit} (\texttt{lqubit}).
To obtain propositional instances, one substitutes all objects with corresponding types in the predicates.
This process is called grounding. It translates the PDDL description to a (potentially very large) propositional problem.

We use the \texttt{map\_initial} action to map a logical qubit \texttt{?l} to a physical qubit \texttt{?p}.
One can apply the action only if \texttt{?l} and \texttt{?p} are both not mapped.
Once the action is applied, \texttt{?l} is mapped to \texttt{?p}.
\begin{lstlisting}[caption={predicates and action for initial mapping},label={lst:initialmappingdomain},
language=stripspddl,basicstyle=\small,mathescape]
(:types lqubit pqubit)
(:predicates (mapped ?l - lqubit ?p - pqubit)
             (mapped_lq ?l - lqubit)
             (mapped_pq ?p - pqubit))
(:action map_initial
 :parameters (?l - lqubit ?p - pqubit)
 :precondition (and
   (not(mapped_lq ?l)) (not(mapped_pq ?p)))
 :effect (and (mapped ?l ?p)
   (mapped_lq ?l) (mapped_pq ?p)))
\end{lstlisting}

\paragraph*{Valid Gate Dependencies}
To enforce gate dependencies in this model, we use the layers as generated by Qiskit.%
\footnote{This may introduce spurious dependencies, when independent gates end up in different layers.
This sub-optimality is solved in our Local model.}
As long as gates are executed layer by layer, their dependencies will be respected. 
When modelling layout synthesis as planning, we only consider the layers with CNOT gates.
To represent layers in our first planning model, we introduce objects of type \texttt{depth},
representing discrete depths, i.e., one object for each layer that contains a CNOT gate.
%
In the DAG of the adder example in Figure \ref{fig:addercircuit},
only the layers at depths $d2,d3,d4,d5,d6,d8,d{10}$ contain CNOT gates.

To ensure the circuit is executed layer by layer, we maintain a predicate \texttt{current\_depth} 
to specify the current layer, and use a static predicate \texttt{next\_depth} to relate the current and the next layer.
We use the action \texttt{move\_depth} to move from one layer to another layer in a plan.
Every valid plan has exactly $k$ \texttt{move\_depth} actions, where $k$ is the number of layers in the DAG with at least one CNOT gate.

Using layers provides two advantages over placing each instruction in its own layer,
following the OPENQASM input order: (1) There are fewer dependencies, potentially reducing
the number of swaps; (2) There are fewer \texttt{move\_depth} actions in the plan, which
typically decreases the solving time.


\begin{lstlisting}[caption={predicates and action for valid gate dependencies},label={lst:validdependenciesdomain},
language=stripspddl,basicstyle=\small,mathescape]
(:types depth)
(:predicates (current_depth ?d - depth)
             (next_depth ?d1 ?d2 - depth))
(:action move_depth
 :parameters (?d1 ?d2 - depth) 
 :precondition (and (current_depth ?d1)
   (next_depth ?d1 ?d2))
 :effect (and (not(current_depth ?d1))
   (current_depth ?d2)))
\end{lstlisting}

\paragraph*{Valid CNOT and SWAP gates}
For the final constraint, we need to specify when CNOT and SWAP gates can be applied.
To model that a CNOT gate must be applied on specific qubits at a certain depth, 
we use predicate \texttt{rcnot} (required-cnot) in Listing \ref{lst:validcnotsdomain}.
Predicate \texttt{(rcnot ?l1 ?l2 ?d)} indicates that a CNOT gate must
still be applied between two logical qubits \texttt{?l1}, \texttt{?l2} at the layer \texttt{?d}.
Here \texttt{?l1, ?l2, ?d} are parameters that can be substituted with objects of the corresponding types 
to generate (grounded) propositions.

To apply a CNOT gate, the corresponding logical qubits must be mapped to some physical qubits;
these physical qubits must be connected; and the current depth must match a required CNOT gate.
The action \texttt{apply\_cnot} ensures that all the above conditions are satisfied.
The effect of this action removes one required CNOT operation (\texttt{rcnot}).

Finally, we model the action \texttt{swap} to allow swapping of logical qubits when required.
We allow two types of swaps:
\begin{itemize}
  \item Swapping two mapped logical qubits.
  \item Swapping a mapped logical qubit with an ancillary qubit.
\end{itemize}
Note that we call both actions \texttt{swap}, although their parameters and conditions are different.
In the first case, similar to a CNOT gate, both logical qubits must be mapped to connected physical qubits.
The effect will update the mapping from logical to physical qubits.
In the second case, a logical qubit \texttt{?l1} must be mapped to a physical qubit \texttt{?p1},
which should be connected to a free (not mapped) physical qubit \texttt{?p2}.
Once swapped, \texttt{?l1} is now mapped to \texttt{?p2}, and \texttt{?p1} becomes free again (not mapped).
If the coupling graph is \emph{not} bidirectional, we need a symmetric version of the ancillary swap as well
(see Appendix~\ref{app:localinitial} for full details).

\begin{lstlisting}[caption={predicates and actions for valid CNOT and SWAP gates},label={lst:validcnotsdomain},
language=stripspddl,basicstyle=\small,mathescape]
(:predicates
   (rcnot ?l1 ?l2 - lqubit ?d - depth)
   (connected ?p1 ?p2 - pqubit))

(:action apply_cnot
 :parameters (?l1 ?l2 - lqubit
   ?p1 ?p2 - pqubit ?d - depth)
 :precondition (and (connected ?p1 ?p2)
   (mapped ?l1 ?p1) (mapped ?l2 ?p2)
   (rcnot ?l1 ?l2 ?d) (current_depth ?d))
 :effect (and (not(rcnot ?l1 ?l2 ?d))))

(:action swap ;; with another mapped qubit
 :parameters (?l1 ?l2 - lqubit
              ?p1 ?p2 - pqubit)
 :precondition (and (connected ?p1 ?p2)
   (mapped ?l1 ?p1) (mapped ?l2 ?p2) )
 :effect (and 
   (not(mapped ?l1 ?p1)) (mapped ?l1 ?p2)
   (not(mapped ?l2 ?p2)) (mapped ?l2 ?p1)))

(:action swap ;; with an ancillary qubit
 :parameters (?l1 - lqubit ?p1 ?p2 - pqubit)
 :precondition (and (connected ?p1 ?p2)
   (mapped ?l1 ?p1) (not(mapped_pq ?p2)) )
 :effect (and 
   (not(mapped ?l1 ?p1)) (mapped ?l1 ?p2) 
   (not(mapped_pq ?p1)) (mapped_pq ?p2)))
  
\end{lstlisting}

\subsubsection{Problem Specification}
In the problem file, we need to specify the initial and goal states of the planning problem.
For the adder example, the initial depth is the first layer with a CNOT, \texttt{d2}.
We specify the coupling map of the architecture using connected predicates.
For CNOT mapping, we specify all the required CNOT gates at specific depths.
Finally, the \texttt{next\_depth} predicate specifies a total ordering of the layers.
Note that the initial state specification is complete, i.e., all propositions not specified are negated.
In case of the goal specification, one only needs to specify the propositions that must be achieved.
Propositions that are not specified are open, resulting in more than one goal state.
In our model, we specify that all logical qubits are mapped and all required CNOT gates are 
performed (i.e., each \texttt{rcnot} is negated).
Listings \ref{lst:adderinitialsnippets} and \ref{lst:goalstatesnippets}
present snippets of initial and goal specification of our adder example.


\begin{lstlisting}[caption={Adder initial state snippets},label={lst:adderinitialsnippets},
   language=stripspddl,basicstyle=\small,mathescape]
(:init
 (current_depth d2)
 (connected p1 p0) (connected p1 p2) ...
 (next_depth d2 d3) (next_depth d3 d4) ...
 (rcnot l2 l3 d2) (rcnot l0 l1 d3) ... )
\end{lstlisting}

\begin{lstlisting}[caption={Adder goal state snippets},label={lst:goalstatesnippets},
language=stripspddl,basicstyle=\small,mathescape]
(:goal
 (and
  (mapped_lq l0) ... (mapped_lq l3)
  (not (rcnot l2 l3 d2)) ... ))
\end{lstlisting}

\subsection{Local Dependency Based Model}
\label{subsec:localdependencybased}

In the Global model, using explicit actions for moving to the next depth results in almost doubling the plan length in the worst case.
At least, the plan length increases by the number of layers in the circuit with CNOT gates.
It is well known that increased plan length can result in increased solving time.
Additionally, we need initial mapping actions for mapping logical qubits to physical qubits.
This further increases the plan length. These initial mapping steps also create a large search space,
which is not a priori pruned by the needs of the logical circuit.
Moreover, since the layers introduce spurious dependencies, the resulting number
of SWAPs could be suboptimal.

To overcome these problems, we make three improvements to our Global encoding.
First, we drop the layers and handle local dependencies directly when applying CNOT gates.
Second, we integrate the initial mapping into the corresponding CNOT gates.
Third, as a final optimization, we partially ground the gates in the \texttt{apply\_cnot} 
actions to reduce the number of parameters.
We now describe these improvements in detail. 

\subsubsection{Avoiding Layers (Local/Initial)}
\label{subsubsec:localinitial}

In all our actions and predicates, we drop the depth parameters, and we drop the action \texttt{next\_depth}.
Instead, we will specify gate dependencies locally. Recall that during planning we only consider
the binary CNOT gates, and we completely ignore the unary gates.
Then a CNOT gate $C_1$ depends on at most two preceding CNOT gates $C_2$ and $C_3$.
A gate could also directly depend on a logical input qubit.
As long as $C_2$ and $C_3$ are executed before $C_1$, we preserve the correctness.

To model local dependencies (cf.\ Listing~\ref{lst:localinitialpredicates}), 
we introduce a static predicate \texttt{(cnot ?l1 ?l2 ?g1 ?g2 ?g3)},
which indicates that the logical circuit contains a gate \texttt{?g1} on logical qubits
\texttt{?l1} and \texttt{?l2}, that depends on previous gates \texttt{?g2} and \texttt{?g3}.
To avoid exceptions for gates that directly depend on the input qubits, we view \texttt{lqubit}
as a subtype of \texttt{gate}. We introduce predicate \texttt{done} to keep track of the
CNOT gates that have already been applied. Using \texttt{done}, we don't require
\texttt{mapped\_lq} anymore, and we rename \texttt{mapped\_pq} to \texttt{occupied}.
The predicates \texttt{mapped} and \texttt{connected} are similar as before.

\begin{lstlisting}[caption={Local/Initial Domain: types and predicates},label={lst:localinitialpredicates},
  language=stripspddl,basicstyle=\small,mathescape]
(:types 
  pqubit gate - object
  lqubit - gate) 
(:predicates
  (cnot ?l1 ?l2 - lqubit ?g0 ?g1 ?g2 - gate)   
  (done ?g - gate)                        
  (mapped ?l - lqubit ?p - pqubit)
  (occupied ?p - pqubit)                        
  (connected ?p1 ?p2 - pqubit))
\end{lstlisting}

The actions \texttt{map\_initial} and \texttt{apply\_cnot} are adapted as in
Listing~\ref{lst:localinitialactions}. Note that an \texttt{lqubit} becomes done
when it is mapped, while a CNOT gate becomes done when it is applied.
Also note that in the latter case, we require that the dependent CNOT
gates (or logical input qubits) are already done.
The \texttt{swap}-actions are similar to the global encoding
in Listing~\ref{lst:validcnotsdomain}.

\begin{lstlisting}[caption={Local/Initial Domain: actions \texttt{map\_initial} and \texttt{apply\_cnot}},
  label={lst:localinitialactions}, language=stripspddl,basicstyle=\small,mathescape]
(:action map_initial
 :parameters (?l - lqubit ?p - pqubit)
 :precondition 
   (and (not(done ?l)) (not(occupied ?p)))
 :effect (and (done ?l) 
   (mapped ?l ?p) (occupied ?p)))

(:action apply_cnot
 :parameters (?l1 ?l2 - lqubit 
              ?p1 ?p2 - pqubit 
              ?g0 ?g1 ?g2 - gate)
 :precondition (and
     (cnot ?l1 ?l2 ?g0 ?g1 ?g2)
     (connected ?p1 ?p2)
     (mapped ?l1 ?p1) (mapped ?l2 ?p2)
     (done ?g1) (done ?g2) (not(done ?g0)))
 :effect (and (done ?g0)))
\end{lstlisting}

\subsubsection{Integrating the Initial Mapping (Lifted/Compact)}
\label{subsubsec:iftedcompact}

Recall that not all CNOT gates have two CNOT dependencies; 
for example gate $g4$ in Figure \ref{fig:addercircuit} depends
directly on two logical (input) qubits \texttt{l2} and \texttt{l3}.
To avoid explicit initial mapping actions, we integrate the mapping of these input qubits 
directly into the CNOT actions.

In this lifted compact model, we drop the \texttt{map\_initial} action entirely. 
Instead, we split the \texttt{apply\_cnot} action in four cases, depending on whether 
this gate depends on previous gates or on logical input qubits. In the latter case,
we immediately map the logical qubits to any unmapped, connected physical qubits.
In Listing~\ref{lst:liftedcompactactions} we show two cases, where gate \texttt{?g0}
depends on two gates, or on two inputs, respectively. 
Note that in the latter case we can drop the test for \texttt{(done ?l1)} and \texttt{(done ?l2)}, 
since in this model a logical qubit can only be the input to a single CNOT gate 
(ignoring intermediate unary gates). The full model also includes mixed
cases for \texttt{apply\_cnot\_gate\_input} and \texttt{apply\_cnot\_input\_gate}.

\begin{lstlisting}[caption={Lifted/Compact Domain: \texttt{apply\_cnot} actions},
  label={lst:liftedcompactactions}, language=stripspddl,basicstyle=\small,mathescape]
(:action apply_cnot_gate_gate
 :parameters (?l1 ?l2 - lqubit 
             ?p1 ?p2 - pqubit 
             ?g0 ?g1 ?g2 - gate)
 :precondition (and
     (cnot ?l1 ?l2 ?g0 ?g1 ?g2)
     (connected ?p1 ?p2)
     (mapped ?l1 ?p1) (mapped ?l2 ?p2)
     (done ?g1) (done ?g2) (not (done ?g0)))
 :effect (and (done ?g0)))

(:action apply_cnot_input_input
 :parameters (?l1 ?l2 - lqubit 
              ?p1 ?p2 - pqubit ?g0 - gate)
 :precondition (and
     (cnot ?l1 ?l2 ?g0 ?l1 ?l2)
     (connected ?p1 ?p2)
     (not (occupied ?p1)) (not (occupied ?p2))
     (not (done ?g0))
 )
 :effect (and (done ?g0)
     (mapped ?l1 ?p1) (occupied ?p1)
     (mapped ?l2 ?p2) (occupied ?p2)))
\end{lstlisting}

\subsubsection{Grounding the Gate Names (Local/Compact)}
\label{subsubsec:localcompact}

In both local models, the \texttt{cnot}-predicate and the \texttt{apply\_cnot} action
depend on many parameters. This could slow down the planning tools, since they have
to instantiate these actions for all possible object combinations.
Note that for each gate, the dependencies and the logical qubits they are applied to 
are known at compile time. Therefore, as a final optimization step, in our Local/Compact
model we specialize the \texttt{apply\_cnot} actions in the domain specification for each gate.

The domain file now depends on the logical circuit instance, since it refers to its gates and qubits.
This requires the introduction of
constants with the new type \texttt{gateid} (one for each CNOT gate), as well as
constants for the logical qubits. The mapped physical qubits are not known at compile time,
but will be determined by the planner tool. 

For example, the CNOT gate $g11$ in the adder circuit in Fig.~\ref{fig:addercircuit}
is represented by the constant \texttt{g11}, and it is handled
by the action \texttt{apply\_cnot\_g11}.
It is applied to the logical qubits \texttt{l3} and \texttt{l0}. 
This gate depends on two CNOT gates \texttt{g9} and \texttt{g10},
so the precondition also requires
that the dependent CNOT gates \texttt{g9} and \texttt{g10} are already done.
The complete action specification for this gate becomes:

\begin{lstlisting}[caption={A partially grounded CNOT action for the adder circuit},label={lst:examplecnotdomain},
language=stripspddl,basicstyle=\small,mathescape]
(:types gateid lqubit pqubit)
(:constants g4 g9 g10 g11 ... g22 - gateid 
            l0 l1 l2 l3 - lqubit)
(:action apply_cnot_g11
 :parameters (?p1 ?p2 - pqubit)
 :precondition (and (not (done g11))
   (done g9) (done g10)
   (mapped l3 ?p1) (mapped l0 ?p2)
   (connected ?p1 ?p2))
 :effect (and (done g11)))
\end{lstlisting}

The CNOT gate $g4$ depends on two inputs, so the corresponding action \texttt{apply\_cnot\_g4} 
will integrate two initial mappings as follows:
\begin{lstlisting}[caption={CNOT gate with integrated initial mapping},label={lst:examplecnotdomaing4},
  language=stripspddl,basicstyle=\small,mathescape]
(:action apply_cnot_g4
 :parameters (?p1 ?p2 - pqubit)
 :precondition (and (not (done g4))
   (not(occupied ?p1)) (not(occupied ?p2))
   (connected ?p1 ?p2))
 :effect (and (done g4)
   (mapped l2 ?p1) (occupied ?p1)
   (mapped l3 ?p2) (occupied ?p2)))
\end{lstlisting}
A similar integration is also applied to CNOT gates which depend only on one other CNOT gate.
We provide the full specification for the Adder example in Appendix \ref{sec:appadderpddl}.

The Local Compact encoding avoids spurious dependencies due to using fixed levels. It also results
in shorter plans, since we avoid the \texttt{next\_depth} and \texttt{map\_initial} actions.
Searching for shorter plans should be more efficient. We also expect that the efficiency
is further improved since the actions have fewer parameters (due to grounding) and logical
qubits will only be mapped to connected physical qubits, as required by the initial CNOT gates.
The following section compares the performance of the ``Global'', the (lifted) ``Local Initial''
and the (grounded) ``Local Compact'' encodings experimentally.

\section{Experimental Evaluation}
\label{sec:experimentalevaluation}

\subsection{Experiment Design}
\label{subsec:design}

We have implemented the translation of quantum circuits into PDDL instances in our tool Q-Synth (Quantum Synthesizer).%
\footnote{Q-Synth is available open source at \url{https://github.com/irfansha/Q-Synth/}.}
For an experimental evaluation, we consider a standard set of 16 benchmark circuits from \cite{DBLP:conf/iccad/TanC20},
and we map them onto IBM platforms Tenerife (5 qubits) and Melbourne (14 qubits).
The first experiment maps onto platform Tenerife (also called IBM-X2). 
We consider those 6 out of 16 circuits that use at most the 5
physical qubits available in Tenerife.
For this experiment, we give a 300 seconds time limit and a 16-GB memory limit.
The second experiment maps onto platform Melbourne. We try all 13 out of 16 circuits that
use at most 14 qubits. 
We give a 3-hour time limit and a 48-GB memory limit for all 13 instances.
In both cases, we measure and report the CPU time taken by the whole tool chain (including
parsing, encoding, searching for an optimal plan, and extracting and validating the mapped circuits).

Q-Synth requires a planning tool to solve the generated PDDL instances. 
We consider the Fast Downward planner (FD) with two planner configurations, 
the Big Joint Optimal Landmarks Planner (BJOLP)~\cite{domshlak2011bjolp} and 
Merge-and-Shrink (MS)~\cite{sievers2018fast}, 
and the SAT based planner Madagascar (M)~\cite{Rintanen2014MadagascarS}.
For both experiments, we apply FD+BJOLP on the Global models (column G-bj)
and Lifted Initial models (column LI-bj).
On the Local Compact models,
we report results from all three planners FD with BJOLP (L-bj), FD with MS (L-ms), and Madagascar (L-M).
Note that Madagascar is a satisfying planner. We increment the plan length by 1 until we find a plan.
This ensures optimality of the plans generated by Madagascar.

We compare our results with three existing tools. First 
with Qiskit's heuristic approach SABRE \cite{DBLP:conf/asplos/LiDX19}.
We use the first 100 seeds for SABRE Layout and take the minimum number of swaps produced by any seed.
In every instance, this approach with 100 runs takes approximately 2 minutes.

Then we compare with two leading exact approaches QMAP \cite{DBLP:conf/aspdac/BurgholzerSW22, 10.1145/3593594} and OLSQ \cite{DBLP:conf/iccad/TanC20}.
Although there are heuristic versions of both tools, we only consider the exact versions of them.
We use the latest version OLSQ \texttt{0.0.4.1} installed using pip.
Note that OLSQ only uses bidirectional coupling graphs, so for sake of uniformity we change
the coupling maps of Tenerife and Melbourne for all experiments.
For QMAP, we use the latest version \texttt{mqt.qmap 2.1.3}. 
By default, it uses subset optimization (QMAP-S) as in~\cite{DBLP:conf/aspdac/BurgholzerSW22}.
However, QMAP-S was shown to be suboptimal for certain architectures~\cite{10.1145/3593594}.
Essentially, QMAP-S maps a given circuit with $k$ logical qubits to a subset of $k$ physical qubits.
Instead, QMAP-SA uses optimal subarchitectures that also allow ancillary qubits.
We compare Q-Synth with the exact approaches QMAP (without subarchitectures) and QMAP-SA (with optimal subarchitectures) 
as described in~\cite{10.1145/3593594}.
We also experiment with QMAP-S on the Melbourne platform, but for a fair comparison, in this case we also disable
ancillary swaps in Q-Synth, run with planner FD+BJOLP (L-bj-na).

\subsection{Validation}
\label{subsec:validation}

Validating if our mapped circuits are optimal and correct is non-trivial.
We use the following four measures to ensure correctness and optimality.
\begin{itemize}
\item When a circuit is mapped, we try to recover the original circuit by reversing the swaps and using the reverse initial mapping.
The resultant circuit must be exactly the same as the original circuit, which we compare using Qiskit.
\item We check if all the 2-qubit gates are applied on the connected qubits. One could alternatively use the checkmap function from QISKIT.
\item We compare the equivalence of the mapped circuit with the original circuit with 
at least 1 million simulations in Qiskit.
\item Finally, for optimality we cross compare with the results obtained from other exact tools.
\end{itemize}
Note that with SAT based planners, one could extract certificates when refuting an instance.
This provides provable lower bounds on the number of swaps.

\subsection{Results}
\label{subsec:results}

In Table \ref{table:tenerifedata} (platform Tenerife), Q-Synth with both models and all planner configurations, 
and both existing tools QMAP and OLSQ, solve all the instances.
Overall, for the benchmarks on this small platform, Q-Synth performs similar to QMAP, whereas OLSQ is considerably slower, 
especially on the larger instances.
All exact approaches compute the minimum number of swaps required for \texttt{mod5mils\_65} as 2.
On the other hand, SABRE returns 3 swap insertions, which is non-optimal.

Table \ref{table:melbournedata} presents the results on platform Melbourne with 14 qubits. 
Our Q-Synth with Local Initial models performs much better than with Global models.
Further, Q-Synth with Local Compact models performs much better than with Local Initial models, especially on 9 qubit instances.
Q-Synth solves most instances with planner FD with MS (L-ms) or BJOLP (L-bj): 11 out of 13 within the time and memory limits;
with Madagascar it solves 8 out of 13 instances (L-M); surprisingly L-M times out on one 5-qubit instance.
Even our Global encoding with FD+BJOLP (G-bj) solves 6 out of 13 instances, which is more than previous exact methods. 
Other planning tools on our Global models did not perform that well (timings are not reported here).
Of the previous exact approaches, OLSQ solves only 3 out of 13 instances in total, while QMAP-SA solves 2 out of 13.
Q-Synth is 2 orders of magnitude faster than OLSQ on the \texttt{or} instance.
On the other hand, QMAP (without subarchitectures) runs out of memory for all instances.
In case of solving without ancillary qubits, QMAP-S solves 7 instances, including one 7-qubit instance.
Q-Synth with Local models without ancillary swaps (L-bj-na) still solves the same 11 instances 
optimally (although this combination is suboptimal in general).
It runs up to ten times faster than L-bj (with ancillary swaps).
So considering ancillary swaps can increase the solving time considerably.

We don't know the minimal number of SWAP gates required for the largest
benchmarks, since none of the exact tools could solve them. 
On the other hand, with SABRE we could compute an upper bound on this number. 
One might also find such plans by using QMAP and OLSQ in satisfying 
mode, or running the planning tools on our encoding in satisfying mode,
but this has not been the focus of the research reported here.
Note that for 5 instances, SABRE gives non-optimal swap insertions (this phenomenon was also observed in \cite{DBLP:journals/tc/TanC21}).
For example, for \texttt{4gt13\_92}, SABRE gives 3 additional swaps, which can be implemented by 9 additional CNOT gates.
Within 2 minutes, our approach L-ms computed the minimum number of swaps needed, which then also provides a guaranteed lower bound.
This demonstrates the feasibility of classical planning for exact optimal layout synthesis of quantum circuits of moderate size.

\begin{table*}[htbp]
  \caption{Platform Tenerife or IBM-QX2 (5 qubits), q: Number of logical qubits,
  c: Number of CNOT gates, +s: Number of swaps added,\\ *: non-optimal count.
  We specify the time taken by all exact tools in seconds.}
  \centering
  \begin{tabular}{l|lll|lllll|ccc|c}
  \hline
  & & & &  \multicolumn{5}{c}{Our tool Q-Synth (exact)} & \multicolumn{3}{c}{Previous Exact Tools} & SABRE \\
  \hline
  Circuit      & Q & C  & +S         & G-bj & LI-bj & L-ms & L-bj & L-M & QMAP & QMAP-SA & OLSQ & +S \\  \hline
  or           & 3 & 6  & 0          & 4.9  & 4.3   & 4.1  & 4.0  & 3.9 & 4.0  & 3.9     & 6.7  & 0 \\
  adder        & 4 & 10 & 1          & 4.4  & 4.2   & 4.4  & 4.0  & 4.7 & 3.9* & 4.1*    & 40.3 & 1 \\
  qaoa5        & 5 & 8  & 0          & 3.9  & 4.1   & 4.6  & 4.2  & 4.0 & 4.0  & 3.9     & 12.6 & 0 \\
  4mod5-v1\_22 & 5 & 11 & 1          & 4.7  & 4.1   & 4.7  & 4.1  & 4.1 & 4.0  & 3.9     & 24.2 & 1 \\
  mod5mils\_65 & 5 & 16 & \textbf{2} & 4.2  & 4.0   & 5.3  & 4.3  & 5.4 & 4.0  & 4.0     & 107  & 3 \\
  4gt13\_92    & 5 & 30 & 0          & 4.3  & 4.2   & 6.4  & 4.1  & 5.0 & 4.1  & 4.4     & 136  & 0 \\
  \hline
  \end{tabular}
  \label{table:tenerifedata}
\end{table*}

\begin{table*}[htbp]
  \caption{Platform Melbourne (14 qubits), q: Number of logical qubits,
  c: Number of CNOT gates, +s: Number of swaps added,\\
  TO/MO: Time/Memory Out, *: suboptimal count.
  We specify the time taken by all exact tools in seconds.}
  \centering
  \begin{tabular}{l|lll|lllll|ccc|cc|c}
  \hline
  & & & &  \multicolumn{5}{c}{Our tool Q-Synth (exact)} & \multicolumn{3}{c}{Previous Exact Tools} & \multicolumn{2}{c}{Without ancillary qubits} & SABRE \\
  \hline
  Circuit         & Q  &C   &+S           & G-bj & LI-bj         & L-ms           & L-bj          & L-M           & QMAP & QMAP-SA & OLSQ & L-bj-na & QMAP-S & +S \\
  \hline
  or              & 3  & 6  & 2           & 4.9  &  \textbf{4.1} & 7.6            & \textbf{4.1}  & 4.3           & MO   & 5.5     & 517  & 4.2     &  2.8   & 2 \\
  adder           & 4  & 10 & 0           & 5.5  &  4.5          & 15.1           & \textbf{4.2}  & 4.6           & MO   & 6.5     & 30.5 & 4.3     &  2.4   & 0 \\
  qaoa5           & 5  & 8  & 0           & 5.0  &  4.5          & 23.7           & 4.4           & \textbf{4.1}  & MO   & TO      & 39.5 & 4.2     &  2.6   & 0 \\
  4mod5-v1\_22    & 5  & 11 & \textbf{3}  & 115 &  29.2         & 25.6           & \textbf{5.1}  & 7.4           & MO   & TO      & TO   & 4.7     &  10.1  & 4 \\
  mod5mils\_65    & 5  & 16 & 6           & 1825   &  57.8         & 33.7           & \textbf{7.0}  & 16.3          & MO   & TO      & TO   & 4.6     &  18.6  & 6\\
  4gt13\_92       & 5  & 30 & \textbf{10} & TO   &  280          & \textbf{85.8}  & 121           & TO            & MO   & TO      & TO   & 11.5    &  183*  & 13 \\
  tof\_4          & 7  & 22 & 1           & 5831   &  2716         & 125            & \textbf{4.7}  & 12.3          & MO   & TO      & TO   & 5.1     &  6734  & 1 \\
  barenco\_tof\_4 & 7  & 34 & \textbf{5}  & TO   &  10643        & 184            & 29.9          & \textbf{27.7} & MO   & TO      & TO   & 8.4     &   TO   & 6 \\
  tof\_5          & 9  & 30 & 1           & TO   &  TO           & 386            & \textbf{5.0}  & 449           & MO   & MO      & TO   & 4.7     &   TO   & 1\\
  mod\_mult\_55   & 9  & 40 & \textbf{7}  & TO   &  TO           & \textbf{2316}  & 7710          & TO            & MO   & MO      & TO   & 761     &   TO   & 8 \\
  barenco\_tof\_5 & 9  & 50 & \textbf{6}  & TO   &  TO           & 634            & \textbf{187}  & TO            & MO   & MO      & TO   & 23.2    &   TO   & 7\\
  vbe\_adder\_3   & 10 & 50 & -           & TO   &  TO           & TO             & TO            & TO            & MO   & MO      & TO   & TO      &   TO   & 8 \\
  rc\_adder\_6    & 14 & 71 & -           & TO   &  TO           & TO             & TO            & TO            & MO   & MO      & TO   & TO      &   MO   & 12 \\
  \hline
  \multicolumn{4}{l|}{Total number of instances solved:} &
  6 & 8 & 11 & 11 & 8 & 0 & 2 & 3 & 11 & 7 & 13 \\
  \hline
  \end{tabular}
  \label{table:melbournedata}
\end{table*}

\section{Comparison with Related Work}
\label{sec:relatedwork}

\subsection*{Comparison to QMAP and OLSQ}
\label{subec:discussion}

Our tool Q-Synth with the Local model outperforms QMAP significantly: Q-Synth solves 9 instances uniquely.
The difference is clear when the number of qubits is more than 4: all the instances timed out for QMAP after 3 hours.
Note that Q-Synth solves all three 9-qubit instances (two even within 4 minutes).
QMAP uses an exponential number of variables in the number of qubits to represent the permutations.
We conjecture that is the reason for poor performance on the 14 qubit platform.
Even with sub-architecture optimization, scaling to 14 qubit platform seems difficult, and
we expect the same challenge for even larger platforms.
A similar bottleneck for QMAP is reported in \cite{DBLP:conf/micro/MolaviXDPTA22} with platform Tokyo (20 qubits).
The number of SWAPs in QMAP depends on the order of the input circuit, which can result in suboptimal solutions.
For example, QMAP and QMAP-SA give 2 additional swaps for the \texttt{adder} instance, instead of 1 optimal swap
(${}^*$ in Table~\ref{table:tenerifedata}).

Even our Global encoding outperforms QMAP and solves three 5-qubit instances and one 7-qubit instance.
Avoiding the layers in Local Initial already improves the performance up to an order of magnitude.
Integrating the initial maps with CNOT actions improves the performance even further.
This shows the strength of planning based approaches, and the effect of modelling the problem
(Global vs Local) on performance.

OLSQ (run in exact mode for optimal solutions) scales poorly with the number of qubits and CNOT gates compared to Q-Synth.
Between OLSQ and QMAP-SA, QMAP-SA performs better on a 5-qubit platform whereas OLSQ performs better on a 14-qubit platform.
These results are consistent with the exponential dependency of QMAP on the target platform.
Note that other approaches could also benefit from the optimal subarchitectures computed by QMAP-SA.
Further, when optimizing for the number of swaps, the solutions from OLSQ are not necessarily optimal.
OLSQ optimizes swap count for each depth iteratively and stops at the first depth where a mapping is found.
This can be suboptimal since larger depth mapped circuits are not explored.
In our experiments, the swap counts generated by OLSQ were optimal.

Both tools and also \texttt{satmap} by \cite{DBLP:conf/micro/MolaviXDPTA22} provide heuristic approaches that produce
near optimal solutions and scale much better.
In this paper, we focus only on exact approaches where the solutions are truly optimal in the count of swap insertions.
Proving that there exists no mapping with less SWAP gates is much harder.
For larger circuits, one could apply satisfying planners to our models, to find near-optimal solutions.

\paragraph*{Without Ancillary Qubits}

QMAP-S performs significantly better than QMAP-SA, consistent with the theoretical considerations in~\cite{10.1145/3593594}.
For example, consider the instance \texttt{qaoa5}, QMAP-S tries to map onto a 5-qubit subset of the physical qubits,
while QMAP-SA tries to map onto a 7-qubit subset, i.e., it allows 2 additional ancillary qubits.
Further, QMAP-S imposes a restriction on the number of swaps in front of each gate.
In case of \texttt{qaoa5}, the swap limit is 3, which significantly reduces the solving time.
However, it is not clear to us if this technique preserves the optimality.
Note that, for the same instance, the swap limit with QMAP-SA is 7.
We observed that QMAP-S returns 11 additional swaps for the instance \texttt{4gt13\_92} 
(${}^*$ in Table~\ref{table:melbournedata}), instead of 10 swaps as reported by Q-Synth.
OLSQ in heuristic mode (OLSQ-TB) also gives 10 additional swaps for the same instance, confirming an upper bound of 10 swaps.
Our approach without ancillary swaps still reports the correct minimal swap insertions in all instances.
However, \cite{10.1145/3593594} provides an example where ancillary swaps are essential
to obtain an optimal solution.

\subsection*{Comparison to Temporal Planning}
\label{subec:tempdiscussion}

The authors of \cite{DBLP:conf/ijcai/VenturelliDRF17} proposed to use temporal planners for Layout Synthesis on QAOA problems.
In Section 4 \cite{DBLP:conf/ijcai/VenturelliDRF17}, classical planning is mentioned as a potential alternative
for temporal planning.
However, they use time steps for gates to avoid dependencies, similar to our Global level-based encoding, but without grouping the CNOT gates in layers.
This can almost double the plan lengths, and it blurs the look-ahead information due to lack of dependencies.
In their paper, poor preliminary results were reported with the SAT based planner Madagascar (M/Mp) with parallel plans,
which is consistent with our observations.
Another key issue with their use of temporal planning is the lack of optimality: one can only obtain plans optimal in makespan.
This issue is similar to parallel plans in classical planning: the number of swaps added need not be optimal and
can be worse in practice.

In \cite{DBLP:conf/ecai/DoWOVRF20}, layout synthesis for the QAOA algorithm for graph coloring is split into
two stages: Qubit Initialization (QI) and Routing.
Classical planning is considered for QI, where it is only used for initializing logical qubits on physical qubits.
The addition of SWAP gates and generating actual plans is still handled by temporal planners.
Compared to random initial allocation, using the classical planner FastDownward, improved the makespan in many instances.

\section{Conclusion}
From our experiments, we conclude that classical planning provides
a strong alternative to temporal planning or SMT solving for solving the 
optimal circuit-layout synthesis problem. Moreover, the model using Local
dependencies is superior to the one using Global levels.

\appendix

\subsection{PDDL specification for Adder Circuit -- Local Compact}
\label{sec:appadderpddl}

We provide the detailed domain file and problem file for the adder example
in Fig~\ref{fig:adder}. We map it to the Tenerife platform by using the Local 
Compact encoding.
These files can be generated using our tool
Q-Synth\footnote{\url{https://github.com/irfansha/Q-Synth/releases/tag/Q-Synth-v1.0-ICCAD23}}, 
using the following command, where \texttt{-a1} switches
on the use of ancillary bits, and \texttt{-b1} makes the coupling map bidirectional.

\texttt{q-synth.py -m local -p tenerife -a1 -b1 Benchmarks/adder.qasm}

\begin{lstlisting}[caption={Adder circuit -- Local Compact; domain file},label={lst:adderlocaldomain},
  language=stripspddl,basicstyle=\small,mathescape]
  (define (domain Quantum)
  (:requirements :strips :typing 
                 :negative-preconditions)
  (:types lqubit pqubit gateid - object)
  (:constants g4 g9 g10 g11 g12 g13
              g14 g19 g20 g22  - gateid
              l0 l1 l2 l3 - lqubit)
  (:predicates
    (occupied ?p - pqubit)
    (mapped ?l - lqubit ?p - pqubit)
    (connected ?p1 ?p2 - pqubit)
    (done ?g - gateid))
  (:action swap
   :parameters (?l1 ?l2 - lqubit 
                ?p1 ?p2 - pqubit)
   :precondition (and (connected ?p1 ?p2)
      (mapped ?l1 ?p1) (mapped ?l2 ?p2))
   :effect (and 
      (not (mapped ?l1 ?p1)) (mapped ?l1 ?p2)
      (not (mapped ?l2 ?p2)) (mapped ?l2 ?p1)))
  (:action swap-ancillary1
   :parameters (?l1 - lqubit 
                ?p1 ?p2 - pqubit)
   :precondition (and (connected ?p1 ?p2)
      (mapped ?l1 ?p1) (not (occupied ?p2)))
   :effect (and 
      (not (mapped ?l1 ?p1)) (mapped ?l1 ?p2) 
      (not (occupied ?p1)) (occupied ?p2)))
  (:action swap-ancillary2
   :parameters (?l2 - lqubit ?p1 ?p2 - pqubit)
   :precondition (and (connected ?p1 ?p2)
      (mapped ?l2 ?p2) (not (occupied ?p1)))
   :effect (and 
      (not (mapped ?l2 ?p2)) (mapped ?l2 ?p1) 
      (not (occupied ?p2)) (occupied ?p1)))
  (:action apply_cnot_g4
   :parameters (?p1 ?p2 - pqubit)
   :precondition (and 
      (not (done g4)) (connected ?p1 ?p2)
      (not(occupied ?p1)) (not(occupied ?p2)))
   :effect (and (done g4) 
      (mapped l2 ?p1) (occupied ?p1) 
      (mapped l3 ?p2) (occupied ?p2)))
  (:action apply_cnot_g9
   :parameters (?p1 ?p2 - pqubit)
   :precondition (and 
      (not (done g9)) (connected ?p1 ?p2)
      (not (occupied ?p1)) (not(occupied ?p2)))
   :effect (and (done g9) 
      (mapped l0 ?p1) (occupied ?p1) 
      (mapped l1 ?p2) (occupied ?p2)))
  (:action apply_cnot_g10
   :parameters (?p1 ?p2 - pqubit)
   :precondition (and 
      (not (done g10)) (connected ?p1 ?p2)
      (done g4) (mapped l2 ?p1) 
      (done g4) (mapped l3 ?p2))
   :effect (and (done g10)))
  (:action apply_cnot_g11
   :parameters (?p1 ?p2 - pqubit)
   :precondition (and 
      (not (done g11)) (connected ?p1 ?p2)
      (done g9) (mapped l1 ?p1) 
      (done g10) (mapped l2 ?p2))
   :effect (and (done g11)))
  (:action apply_cnot_g12
   :parameters (?p1 ?p2 - pqubit)
   :precondition (and 
      (not (done g12)) (connected ?p1 ?p2)
      (done g10) (mapped l3 ?p1) 
      (done g9) (mapped l0 ?p2))
   :effect (and (done g12)))
  (:action apply_cnot_g13
   :parameters (?p1 ?p2 - pqubit)
   :precondition (and 
      (not (done g13)) (connected ?p1 ?p2)
      (done g12) (mapped l0 ?p1) 
      (done g11) (mapped l1 ?p2)) 
   :effect (and (done g13)))
  (:action apply_cnot_g14
   :parameters (?p1 ?p2 - pqubit)
   :precondition (and 
      (not (done g14)) (connected ?p1 ?p2)
      (done g11) (mapped l2 ?p1) 
      (done g12) (mapped l3 ?p2))
   :effect (and (done g14)))
  (:action apply_cnot_g19
   :parameters (?p1 ?p2 - pqubit)
   :precondition (and 
      (not (done g19)) (connected ?p1 ?p2)
      (done g13) (mapped l0 ?p1)
      (done g13) (mapped l1 ?p2))
   :effect (and (done g19)))
  (:action apply_cnot_g20
   :parameters (?p1 ?p2 - pqubit)
   :precondition (and 
      (not (done g20)) (connected ?p1 ?p2)
      (done g14) (mapped l2 ?p1) 
      (done g14) (mapped l3 ?p2))
   :effect (and (done g20)))
  (:action apply_cnot_g22
   :parameters (?p1 ?p2 - pqubit)
   :precondition (and 
      (not (done g22)) (connected ?p1 ?p2)
      (done g20) (mapped l3 ?p1) 
      (done g19) (mapped l0 ?p2))
   :effect (and (done g22))))
\end{lstlisting}


\begin{lstlisting}[caption={Adder circuit -- Local Compact; problem file},label={lst:adderlocalproblem},
  language=stripspddl,basicstyle=\small,mathescape]
(define (problem circuit)
(:domain Quantum)
(:objects p0 p1 p2 p3 p4 - pqubit)
(:init
  (connected p1 p0) 
  (connected p0 p1)
  (connected p2 p0) 
  (connected p0 p2)
  (connected p2 p1) 
  (connected p1 p2)
  (connected p3 p2) 
  (connected p2 p3)
  (connected p3 p4) 
  (connected p4 p3)
  (connected p4 p2) 
  (connected p2 p4)
)
(:goal (and
  (done g4)
  (done g9)
  (done g10)
  (done g11)
  (done g12)
  (done g13)
  (done g14)
  (done g19)
  (done g20)
  (done g22))))
  \end{lstlisting}

An optimal plan, generated by FastDownward with BJOLP looks
as follows:
\begin{verbatim}
  (apply_cnot_g9 p0 p1)
  (apply_cnot_g4 p2 p3)
  (apply_cnot_g10 p2 p3)
  (apply_cnot_g11 p1 p2)
  (swap l2 l3 p2 p3)
  (apply_cnot_g12 p2 p0)
  (apply_cnot_g13 p0 p1)
  (apply_cnot_g19 p0 p1)
  (apply_cnot_g14 p3 p2)
  (apply_cnot_g20 p3 p2)
  (apply_cnot_g22 p2 p0)
  
\end{verbatim}

The optimal plan needs 1 swap (cf Fig.~\ref{fig:addermapped}).
The Global encoding can be obtained in a similar manner by specifying
\texttt{-m global} and the lifted Local Initial 
encoding by using \texttt{-m lifted\_initial}.

\subsection{Complete SWAP Actions for Global and Local Initial}
\label{app:localinitial}

\begin{lstlisting}[caption={Complete SWAP actions for Global and Local Initial models},label={lst:actionsLocalInitial},
  language=stripspddl,basicstyle=\small,mathescape]
(:action swap
 :parameters (?l1 ?l2 - lbit ?p1 ?p2 - pbit)
 :precondition (and 
    (connected ?p1 ?p2)
    (mapped ?l1 ?p1) (mapped ?l2 ?p2))
 :effect (and 
    (mapped ?l1 ?p2) (not (mapped ?l1 ?p1))
    (mapped ?l2 ?p1) (not (mapped ?l2 ?p2))))
(:action swap-ancillary1
 :parameters (?l1 - lbit ?p1 ?p2 - pbit)
 :precondition (and 
    (connected ?p1 ?p2)
    (mapped ?l1 ?p1)
    (not (occupied ?p2)))
 :effect (and 
    (not(mapped ?l1 ?p1)) (not(occupied ?p1))
    (mapped ?l1 ?p2) (occupied ?p2)))
(:action swap-ancillary2
 :parameters (?l2 - lbit ?p1 ?p2 - pbit)
 :precondition (and 
    (connected ?p1 ?p2)
    (mapped ?l2 ?p2)
    (not (occupied ?p1)))
 :effect (and 
    (not(mapped ?l2 ?p2)) (not(occupied ?p2))
    (mapped ?l2 ?p1) (occupied ?p1)))
\end{lstlisting}

\section*{Acknowledgement}
We thank L.\ Burgholzer for his generous help with the various versions of the QMAP software.

\medskip
Please cite this paper as:
\begin{verbatim}
@inproceedings{ShaikvdP2023,
  author       = {Irfansha Shaik and 
                  Jaco van de Pol},
  title        = {Optimal Layout Synthesis 
                  for Quantum Circuits 
                  as Classical Planning},
  booktitle    = {{ICCAD'23}},
  address      = {San Diego, CA, USA}},
  organization = {{IEEE/ACM}},
  year         = {2023}
}
\end{verbatim}
  

\IEEEtriggeratref{14}
\bibliographystyle{IEEEtran}
\bibliography{IEEEabrv,references}

\end{document}